\documentclass[%
aps,prl,twocolumn,
superscriptaddress,
]{revtex4-2}

\usepackage{graphicx}
\usepackage{amsfonts}
\usepackage{amssymb}
\usepackage{amsthm}
\usepackage{amsmath}
\usepackage{multirow}

\usepackage{color}
 \usepackage{makecell}
\def\beq{\begin{equation}}
\def\eeq{\end{equation}}
\def\bea{\begin{eqnarray}}
\def\eea{\end{eqnarray}}
\newcommand{\beqs}{\begin{subequations}}
\newcommand{\eeqs}{\end{subequations}}

\newcommand{\cref}[1]{Ref.~\cite{#1}}

\newcommand{\hh}{{\ensuremath{I{\kern-2.6pt h}}}}
\newcommand{\bhh}{{\ensuremath{\bar{I{\kern-2.6pt h}}}}}

\usepackage[colorlinks=true,allcolors=blue]{hyperref}
\usepackage{mathtools}

\usepackage[thinc]{esdiff}

\begin{document}
	
\title{Superconducting Strings in $E_6$}
\author{Rinku Maji}
\affiliation{Cosmology, Gravity and Astroparticle Physics Group, Center for Theoretical Physics of the Universe,  Institute for Basic Science, Daejeon 34126, Republic of Korea}
	\author{Qaisar Shafi}
		\affiliation{Bartol Research Institute, Department of Physics and 
		Astronomy,
		 University of Delaware, Newark, DE 19716, USA}
\begin{abstract}
We discuss the appearance of superconducting strings in $E_6$ grand unification, keeping track of the magnetic monopole flux that precedes the formation of the string flux tube. This flux matching ensures compatibility with the quantum tunneling of a monopole-antimonopole pair on a metastable string. We identify two realistic $E_6$ models with superconducting (metastable) strings that also carry zero modes of the right handed Majorana neutrinos and dark matter particles. Depending on the symmetry breaking scale associated with the strings, the latter could be a  source of observable gravitational waves, intermediate scale dark matter, and the observed baryon asymmetry via leptogenesis. Topologically stable superconducting strings also appear if the $E_6$ symmetry breaking leaves unbroken the $Z_2$ subgroup of $Z_4$, the center of $SO(10)$. The zero modes of the SM fermions are the charge carriers in this case. Finally, the flux matching condition ensures that the Aharonov-Bohm phase change in going around the metastable strings is an integer multiple of $2 \pi$ for all fields.
The fields in the spinorial representation of SO(10) acquire a phase change of $\exp(\pm i\pi)$ if taken around the topologically stable $Z_2$ string.

\end{abstract}

\maketitle
\section{Introduction}
Cosmic strings arise in realistic extensions of the Standard Model with the first \cite{Kibble:1982ae}, and so far the only known example, of topologically stable strings discovered in models based on $SO(10)$ grand unification. On the other hand, metastable strings, which are susceptible to breaking via quantum mechanical tunneling of monopole-antimonopole pairs, are far easier to identify in realistic models. In a pioneering paper \cite{Witten:1984eb}, Witten argued that metastable strings in $E_6$ grand unification can be superconducting. For recent discussions of topological structures in $E_6$ and $SO(10)$ models, see Refs.~\cite{Lazarides:2019xai, Lazarides:2023iim, Afzal:2023kqs, Maji:2024pll, Maji:2025thf, Dunsky:2021tih}. 
Metastable \cite{Buchmuller:2021mbb} and quasistable \cite{Martin:1996ea, Lazarides:2022jgr} cosmic strings have in recent years acquired a fair amount of attention, especially if they are superheavy, with a dimensionless string tension $G\mu \sim 10^{-6}$. It has been shown \cite{NANOGrav:2023hvm, Lazarides:2023ksx, Buchmuller:2023aus, Lazarides:2023rqf, Maji:2023fhv, Antusch:2023zjk, Ahmed:2023rky, Afzal:2023cyp, Ahmed:2023pjl, King:2023wkm, Pallis:2024mip, Maji:2024pll,Maji:2024cwv, Antusch:2024nqg, Maji:2024tzg,Pallis:2024joc, Chitose:2024pmz, Datta:2024bqp, Ahmad:2025dds,Antusch:2025xrs}
that the emission of gravitational waves by such superheavy metastable or quasistable strings provides a plausible explanation of the stochastic gravitational wave detected by the Pulsar Timing Array (PTA) experiments \cite{NANOGrav:2023gor, Antoniadis:2023ott, Reardon:2023gzh, Xu:2023wog}.
Gravitational wave emission from metastable strings carrying zero modes of the right handed Majorana neutrinos has also been recently discussed \cite{Afzal:2023kqs}.

In this paper we provide a unified discussion of metastable and topologically stable superconducting strings in $E_6$ models. We identify two distinct $E_6$ models that yield superconducting metastable strings \cite{Witten:1984eb}. In our discussion we keep track of the magnetic monopoles whose nucleation on the string makes the latter unstable. We refer to it as `flux matching condition'. Following Ref.~\cite{Lazarides:2019xai}, the magnetic flux of the minimally charged magnetic $E_6$ monopole that is associated with the string contains a mixture of $U(1)_\psi$ and $SO(10)$ which is precisely determined by topological considerations. Here we are assuming that $E_6$ is broken via $SO(10) \times U(1)_\psi$ which yields the desired magnetic monopole. Taking this into account allows us to identify the magnetic flux tube subsequently carried by the string. The breaking of $SO(10)\times U(1)_\psi$ symmetry to the SM symmetry can proceed in more than one way that we briefly describe. In particular, we identify an $E_6$ model that yields superconducting metastable strings that also carry zero modes of the three right handed  Majorana neutrinos and the dark matter particle (an $SO(10)$ singlet fermion). If the string is relatively light, namely if the $U(1)$ symmetry breaking scale is on the order of 10 TeV or so, this string can also be a source of eV mass sterile neutrinos.  However, if the string is superheavy, which also leads to much heavier dark matter in the model, the gravitational wave emission from the string can be detected by the ongoing and future PTA measurements. 
 
 For completeness, we show how a topologically stable superconducting string appears in $E_6$ if the symmetry breaking to $SU(3)_c \times U(1)_{\rm em}$ also preserves a $Z_2$ gauge symmetry, which is contained in the center $Z_4$ of $SO(10)$. Finally, the flux matching condition also ensures compatibility with the Aharonov-Bohm (AB) requirement that the phase change of all matter and Higgs fields upon circling around the metastable strings is an integral multiple of $2\pi$. An exception arises for the fields in the spinorial representation of $SO(10)$ in the presence of a topologically stable $Z_2$ string associated with an unbroken $Z_2$ gauge symmetry.

\section{Superconducting metastable strings in $E_6$}
\label{sec:2}
 The chiral (16) and vector (10) representations of $SO(10)$, along with an $SO(10)$ singlet, reside in the 27 dimensional representation in $E_6$ grand unification.
The decomposition of the $E_6$ representations 27 and 78, under $SO(10)\times U(1)_\psi$, followed by the breaking $SO(10)\to SU(5)\times U(1)_\chi$,  is as follows:
\begin{align}
&E_6\to SO(10)\times U(1)_\psi: \nonumber \\  & 27 = 16(1) + 10(-2) + 1(4), \quad 78\supset 1(0)+45(0). \\
& SO(10) \to SU(5) \times U(1)_\chi: \nonumber \\  & 16 = 10(-1) + \bar{5}(3) + 1(-5) , \ 10 = 5(2)+\overline{5}(-2), \nonumber \\  & 45\supset 1(0)+24(0). \\
& SU(5)\to  SU(3)_c \times SU(2)_L \times U(1)_Y: \nonumber \\  & 10 =  (3,2)(1) + (\bar{3}, 1)(-4) + (1,1)(6) ,
\nonumber \\  & \overline{5} = (\overline{3},1)(2) + (1,2)(-3),\quad 24\supset (1,1)(0).
\end{align}
The $U(1)$ generators are normalized following Ref.~\cite{Slansky:1981yr}, such that the multiplets have the minimal charges compatible with a period of $2\pi$. With this normalization the electric charge generator and and its orthogonal broken generator are given by
\begin{align}
\label{eq:Q}
Q=\frac{1}{2}T^3_{L}+\frac{Y}{6} ,
\quad
\mathcal{B}=\frac{1}{2}T^3_{L}-\frac{Y}{10}.
\end{align}
In Table~\ref{tab:charges} we summarize the relevant charges, namely $Y$, $\psi$ and $\chi$ of the multiplets in the 27 representation, along with some other charges associated with the generators which play an important role in our analysis.

{\begin{table*}[htbp]
\begin{center}
\renewcommand{\arraystretch}{1.4}
\begin{tabular}{|c|c|c|c|c|c|c|c|c|c|c|c|c|}
\hline
&\multirow{2}{*}{Fields} & 
\multirow{2}{*}{$Y$} & 
\multirow{2}{*}{$\chi$} & 
\multirow{2}{*}{$\psi$} & 
\multirow{2}{*}{\makecell{$\chi'$}} & 
\multirow{2}{*}{\makecell{$\psi'$}} & 
\multirow{2}{*}{\makecell{$\tfrac{\chi'+\psi'}{4}$\\ $\equiv \tfrac{\chi+\psi}{4}$}}  & \multirow{2}{*}{\makecell{$\mathcal{B}$}} & 
\multirow{2}{*}{\makecell{$\tfrac{\chi'}{4}+\tfrac{\psi'}{20}+\tfrac{Y}{5}$ \\ $\equiv \tfrac{\chi}{5}+\tfrac{Y}{5}$}} & 
\multirow{2}{*}{\makecell{$\tfrac{\psi'}{5}-\tfrac{Y}{5}\equiv$\\$\frac{\psi}{4}+\frac{\chi}{20}-\tfrac{Y}{5}$}} & \multirow{2}{*}{$\tfrac{\chi}{10}-\mathcal{B}$}  & \multirow{2}{*}{$Z_2$} \\
 & & & & & & & & & & & &\\
\hline
{\multirow{11}{*}{\rotatebox[origin=c]{90}{Matter}}}&$u_i$ & $1$ & $-1$ & $1$ & $-1$ & $1$ & $0$ & $\frac{2}{5}$ & $0$ & $0$ & $-\frac{1}{2}$ & $-1$ \\
&$d_i$ & $1$ & $-1$ & $1$ & $-1$ & $1$ & $0$ & $-\frac{3}{5}$ & $0$ & $0$ & $\tfrac{1}{2}$ & $-1$ \\
&$u^c_i$ & $-4$ & $-1$ & $1$ & $-1$ & $1$ & $0$ & $\frac{2}{5}$ & $-1$ & $1$ & $-\tfrac{1}{2}$ & $-1$ \\
&$d^c_i$ & $2$ & $3$ & $1$ & $2$ & $2$ & $1$ & $-\frac{1}{5}$ & $1$ & $0$ & $\tfrac{1}{2}$ & $-1$ \\
& $\nu_i$ & $-3$ & $3$ & $1$ & $2$ & $2$ & $1$ & $\frac{4}{5}$ & $0$ & $1$ & $-\tfrac{1}{2}$  & $-1$ \\
& $e_i$ & $-3$ & $3$ & $1$ & $2$ & $2$ & $1$ & $-\frac{1}{5}$ & $0$ & $1$ & $\tfrac{1}{2}$ & $-1$\\
& $e^c_i$ & $6$ & $-1$ & $1$ & $-1$ & $1$ & $0$ & $-\frac{3}{5}$ & $1$ & $-1$ & $\tfrac{1}{2}$ & $-1$ \\
& $\nu^c_i$ & $0$ & $-5$ & $1$ & $-4$ & $0$ & $-1$ & $0$ & $-1$ & $0$ & $-\tfrac{1}{2}$ & $-1$ \\
& $H_{ui}^+$ & $3$ & $2$ & $-2$ & $2$ & $-2$ & $0$ & $\frac{1}{5}$ & $1$ & $-1$ & 0 & $+1$ \\
& $H_{ui}^0$ & $3$ & $2$ & $-2$ & $2$ & $-2$ & $0$ & $-\frac{4}{5}$ & $1$ & $-1$ & 1 & $+1$ \\
& $H_{di}^0$ & $-3$ & $-2$ & $-2$ & $-1$ & $-3$ & $-1$ & $\frac{4}{5}$ & $-1$ & $0$ & $-1$ & $+1$ \\
& $H_{di}^-$ & $-3$ & $-2$ & $-2$ & $-1$ & $-3$ & $-1$ & $-\frac{1}{5}$ & $-1$ & $0$ & 0 & $+1$ \\
& $D_i$ & $-2$ & $2$ & $-2$ & $2$ & $-2$ & $0$ & $\frac{1}{5}$ & $0$ & $0$ & $0$ & $+1$ \\
& $D^c_i$ & $2$ & $-2$ & $-2$ & $-1$ & $-3$ & $-1$ & $-\frac{1}{5}$ & $0$ & $-1$ & $0$ & $+1$ \\
& $N_i$ & $0$ & $0$ & $4$ & $-1$ & $5$ & $1$ & $0$ & $0$ & $1$ & $0$ & $+1$ \\
\hline
{\multirow{5}{*}{\rotatebox[origin=c]{90}{Higgs}}}& $\nu^c$ & $0$ & $-5$ & $1$ & $-4$ & $0$ & $-1$ & $0$ & $-1$ & $0$ & $-\tfrac{1}{2}$ & $-1$ \\
& $N$ & $0$ & $0$ & $4$ & $-1$ & $5$ & $1$ & $0$ & $0$ & $1$ & 0 & $-1$ \\
& $h_u^+$ & $3$ & $2$ & $-2$ & $2$ & $-2$ & $0$ & $\frac{1}{5}$ & $1$ & $-1$ & $0$ & $+1$ \\
& $h_u^0$ & $3$ & $2$ & $-2$ & $2$ & $-2$ & $0$ & $-\frac{4}{5}$ & $1$ & $-1$ & $1$ & $+1$ \\
& $h_d^0$ & $-3$ & $-2$ & $-2$ & $-1$ & $-3$ & $-1$ & $\frac{4}{5}$ & $-1$ & $0$ & $-1$ & $+1$ \\
& $h_d^-$ & $-3$ & $-2$ & $-2$ & $-1$ & $-3$ & $-1$ & $-\frac{1}{5}$ & $-1$ & $0$ & $0$ & $+1$ \\
& $\nu^c\nu^c$ & $0$ & $-10$ & $2$ & $-8$ & $0$ & $-2$ & $0$ & $-2$ & $0$ & $-1$ & $+1$ \\
\hline
\end{tabular}
\caption{ $U(1)$ charges $Y, \chi, \psi$ of the 27 matter fields in $E_6$, where $i$ stands for the generation index. $N$, $\nu^c$ and $h_u$, $h_d$ denote the relevant Higgs fields in 27, and $\nu^c\nu^c$ denotes the $SU(5)$ singlet in the 126-plet of $\overline{351'}$.
Also shown are quantities that play a role in identifying the superconducting strings. $\mathcal{B}\equiv \tfrac{1}{2}T^3_L - \tfrac{Y}{10}$ is the broken generator orthogonal to $Q$ defined in Eq.~\eqref{eq:Q}, and $\chi'=\tfrac{3\chi-\psi}{4}, \psi'=\tfrac{\chi+5\psi}{4}$ are defined in Eq.~\eqref{eq:chip-psip}. $\tfrac{\chi}{10}-\mathcal{B}$ is the magnetic flux carried by the stable $Z_2$ string. The last column shows the $Z_2$ charges of various fields, where $Z_2\subset Z_4$, the center of $SO(10)$. Note that $\chi =-5(B-L)+\frac{2}{3}Y$, where $B$ and $L$ denote baryon and lepton number respectively.}
\label{tab:charges}
\end{center}
\end{table*}}


We begin with the following breaking of $E_6$:
\begin{align}
E_6 \to SO(10) \times U(1)_\psi.
\end{align}
This breaking was discussed in Ref.~\cite{Witten:1984eb}, and it was argued that the string associated with the $U(1)_\psi$ symmetry is superconducting. In our approach, as we previously mentioned, we ensure in addition, that the $\psi$ monopole flux and the flux carried by the string are compatible with the metastable framework. We note, following Ref.~\cite{Lazarides:2019xai} and Table~\ref{tab:charges}, that the minimally charged $\psi$ monopole carries a flux
$(\psi +\chi)/4$. This flux is compatible with the Dirac quantization condition, as shown for the 27 matter fields in Table \ref{tab:charges}. Note that this flux combination comes about from the non-trivial intersection between the $Z_4$ symmetries in $SO(10)$ and $U(1)_\psi$.

For comparison, let us recall how the topologically stable GUT monopole emerges from the spontaneous breaking of $SO(10)$ to $SU(3)_c \times SU(2)_L \times U(1)_Y$. A $2 \pi$ rotation along $Q$, Eq.~\eqref{eq:Q}, accompanied by suitable color rotation, yields the GUT monopole that carries one unit $2 \pi/e$ of Dirac magnetic charge as well as some color magnetic charge. 

\begin{figure}[htbp]
\begin{center}
\includegraphics[width=0.9\linewidth]{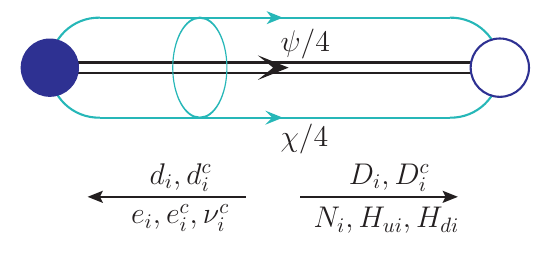}
\caption{$\psi$ monopole-antimonopole pair connected by the superconducting string with magnetic flux $\psi/4+\chi/4$. The right moving and left moving fermion zero modes are also shown. In the presence of an unbroken $Z_2$ symmetry, the lightest $N_i$ is a plausible dark matter candidate.} 
\label{fig:E6_psi_2}
\end{center}
\end{figure}
Returning to the $\psi$ monopole, the breaking of $SO(10)\times U(1)_\psi$ to $SU(3)_c\times SU(2)_L\times U(1)_Y\times U(1)_\chi\times U(1)_\psi$, followed by the breaking of $U(1)_\psi$ and then $U(1)_\chi$, can be accomplished using the $SO(10)$ singlet Higgs field $N$, the $SU(5)$ singlet Higgs field $\nu^c$, and suitable additional Higgs fields as necessary.
This yields a superconducting string in which the supercurrent is carried by the zero modes of  the heavy charged fermions in the 10-plet of $SO(10)$ (right movers), complemented by the SM down quarks and charged leptons including their antiparticles (left movers.) In addition, there will be neutral zero modes of $N_i$ (right movers) and $\nu^c$ (left movers) arising from the dimension five operators $F_i F^\dagger F^\dagger F_i$ with $F$ referring to $N$ or $\nu^c$. The metastable monopole-string system is described in Fig.~\ref{fig:E6_psi_2}. Note that the inner core carries the right movers, while the zero modes of the SM fermions are in the outer core associated with the electroweak flux.

 The phase of the scalar fields $N$ and $\nu^c$ respectively change by $+ 2\pi$ and $-2\pi$ if taken around the string in Fig.~\ref{fig:E6_psi_2}. Note that the $-2\pi$ phase change of $\nu^c$ arises from its interactions with both the $\chi$ and $\psi$ magnetic fluxes. Even though $\chi/4$ is not an integral multiple of the minimal magnetic flux
$\chi/5$, the phase change of $\exp [-2 \pi i (5/4)]$ is accompanied by an additional phase change of $\exp [2 \pi i (1/4)]$ from the interaction of $\nu^c$ with the magnetic flux
$\psi/4$. The overall Aharonov-Bohm phase change is therefore an integral multiple of $2 \pi$. The phase of the up-type electroweak Higgs doublet remains unchanged upon rotation around the string, while the phase of the down-type electroweak doublet changes by $-2\pi$. 

  At this stage it is appropriate to discuss how the superconducting string in Fig.~\ref{fig:E6_psi_2} can be a source of dark matter particles that we alluded to earlier. Let us recall that the breaking of the $SO(10)$ symmetry to $SU(3)_c \times U(1)_{\rm em}$ solely with Higgs fields in tensor representations yields an unbroken $Z_2$ gauge symmetry, which can make a candidate dark matter particle stable. In our case a plausible dark matter candidate can be the lightest $N_i$. However, to realize the $Z_2$ symmetry, we should replace the VEV $\left<\nu^c\right>$ with the VEV $\left<\nu^c \nu^c\right>$ in the $126$-plet of $SO(10)$ which, in turn, is found in the $\overline{351’}$ multiplet of $E_6$. The lightest $N_i$, presumably with an intermediate scale mass, is stable in this case, and its zero mode resides on the superconducting string as shown in Fig.~\ref{fig:E6_psi_2}. Note that the presence of an unbroken $Z_2$ symmetry implies the existence of a topologically stable string in $E_6$ that we shall discuss in Section \ref{sec:3}.

For our second superconducting string which appears in $E_6$, consider the following symmetry breaking:
\begin{align}
\label{eq:breaking-2}
E_6  \to SU(5)\times
U(1)_{\psi'} \to SU(3)_c \times SU(2)_L \times U(1)_Y.
\end{align}
The breaking of $E_6$ to $SU(5)\times U(1)_{\psi’}$ in Eq.~\eqref{eq:breaking-2} is  achieved, among other Higgs scalars, by the VEV of the field $\nu^c$. Here, following Ref.~\cite{Lazarides:2019xai}, we introduce the generator $\psi’$ and its orthogonal counterpart $\chi’$, which are linear combinations of $\psi$ and $\chi$ and defined as follows:
\begin{align}
\label{eq:chip-psip}
\chi'=\frac{3\chi-\psi}{4}, \ \psi'=\frac{\chi+5\psi}{4} .
\end{align}

\begin{figure}[htbp]
\begin{center}
\includegraphics[width=0.9\linewidth]{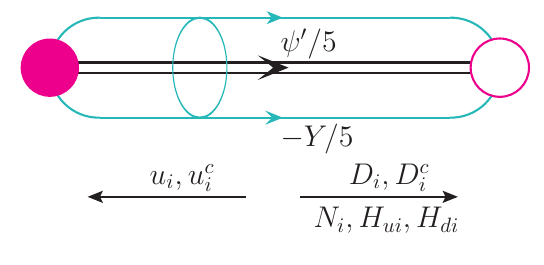}
\caption{ $\psi'$ monopole-antimonopole pair connected by the superconducting string carrying magnetic flux $\psi'/5-Y/5$. The right moving and left moving fermion zero modes are also shown.}
\label{fig:E6_psip_4}
\end{center}
\end{figure}
From Table~\ref{tab:charges} we observe that the $\left<\nu^c\right>$ leaves $U(1)_{\psi’}$ unbroken, while the broken generator is $\chi’$.  The subsequent breaking of $U(1)_{\psi’}$ is achieved with the Higgs field $N$ with a $\psi’$ charge of 5.
From Table~\ref{tab:charges} we conclude that $U(1)_{\psi’}$ and $SU(5)$ share in common a $Z_5$ symmetry. Indeed, the minimal $\psi’$ monopole carries a magnetic flux
$\psi’/5 - Y/5$, which is consistent with the Dirac quantization condition. From Table~\ref{tab:charges}, we can verify that the string is superconducting with the $N$ field responsible for the right movers, and the up-type SM Higgs responsible for the left movers (Fig.~\ref{fig:E6_psip_4}).

\begin{figure}[h!]
\begin{center}
\includegraphics[width=0.95\linewidth]{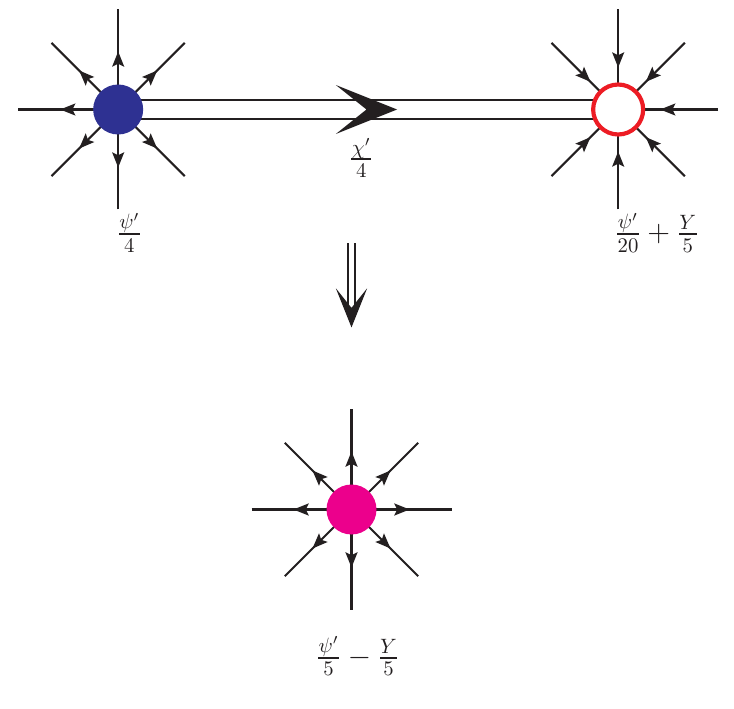}
\caption{ A $\psi$ monopole with flux $\psi'/4+\chi'/4$ ($=\psi/4+\chi/4$), and a $\chi$ antimonopole with flux $\psi'/20 + Y/5$ ($=\chi/5+Y/5$), are connected by $U(1)_{\chi'}$ flux tube as described in the text. They merge together to form a $\psi'$ monopole. This arises from the breaking $E_6 \to SO(10) \times U(1)_\psi \to SU(5)\times
U(1)_\chi \times U(1)_\psi \to SU(5)\times U(1)_{\psi'}$.}
\label{fig:E6_psip_chip}
\end{center}
\end{figure}
To clarify the appearance of the $\psi'$ monopole, consider the symmetry breaking 
\begin{align}
\label{eq:breaking-3}
E_6 & \to SO(10) \times U(1)_\psi \to SU(5)\times
U(1)_\chi \times U(1)_\psi \nonumber \\ &\to SU(5)\times
U(1)_{\psi'}.
\end{align}
The first breaking produces  the $\psi$ monopole  with flux $\psi/4+\chi/4$, which is followed by the $\chi$ monopole in the second breaking which carries a flux of $\chi/5+Y/5$.
In the subsequent breaking of the symmetry $U(1)_\chi \times U(1)_\psi$ to $U(1)_{\psi'}$ with the $SU(5)$ singlet $\nu^c$, the $\psi’$ monopole is realized from the merger of a $\psi$ monopole with a $\chi$ antimonopole, as shown in Fig.~\ref{fig:E6_psip_chip}. [Here we describe the $\psi$ and $\chi$ monopoles in the prime notation $\psi’$, $\chi’$.] 

These $E_6$ models are interesting not only because the string is superconducting, but also because they carry zero modes pertaining to the three right handed neutrinos. Furthermore, if we require the presence of a suitable dark matter particle which, say, is the lightest $SO(10)$ singlet fermion $N_i$, the string also carries the zero modes of dark matter. This is because the $N_i$ matter fields acquire masses from the dimension five couplings
$N_i N_j \bar{N} \bar{N}$, and the lightest one is a plausible dark matter candidate (for recent discussions of dark matter candidates in $E_6$ see \cite{Hebbar:2017fit,Schwichtenberg:2017xhv,Bandyopadhyay:2019rja,Babu:2024ecl,Maji:2024pll,Afzal:2025zii}.)


\section{Topologically stable superconducting string in $E_6$}
\label{sec:3}
We have focused so far on metastable strings in $E_6$ and emphasized the importance of matching the magnetic flux carried by the monopole and string in this setup. In order to implement this scenario, we have utilized the $SU(5)$ singlet Higgs field $\nu^c$ contained in the 27-plet  whose VEV breaks the gauge symmetry $U(1)_\chi$ to $Z_5$.

It was shown in Ref.~\cite{Kibble:1982ae} that if SO(10) is broken to $SU(3)_c \times U(1)_{\rm em}$ using only scalar Higgs in the tensor representations, a discrete $Z_2$ gauge symmetry stays unbroken, which leads to the appearance of topologically stable strings. In $SO(10)$ this means replacing the Higgs 16-plet with a 126-plet which, in $E_6$, can be found in the $351’$ multiplet.

In order to identify the topologically stable string in $E_6$, let us begin by breaking the $E_6$ symmetry to $SO(10)$ using the Higgs field $N$. Next, we employ the 78 dimensional representation to break $SO(10)$ to $SU(5) \times U(1)_\chi$, which produces a chi monopole carrying a magnetic flux $(\chi + Y)/5$. The $\chi$ magnetic flux of this monopole is squeezed into two separate tubes by the VEV of the $SU(5)$ singlet component of the 126-plet (in $351’$). This occurs because the $\chi$ charge of this field is $-10$, which means that the flux tube created from this breaking can only carry a flux $\chi/10$.

\begin{figure}[htbp]
\begin{center}
\includegraphics[width=0.9\linewidth]{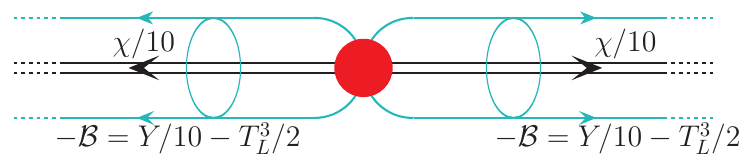}
\caption{Monopole-string system from the breaking $E_6\to SO(10)\to SU(5)\times U(1)_\chi\to SU(3)_c\times SU(2)_L\times U(1)_Y\times Z_2\to SU(3)_c\times U(1)_{\rm em}\times Z_2$. $\mathcal{B}$ denotes the broken electroweak generator orthogonal to $Q$, the electric charge. An open string is stable because of the unbroken $Z_2$ symmetry. The phase change of the matter fields in the spinorial representation of $SO(10)$ is $\exp(\pm i\pi)$ if taken around this superconducting $Z_2$ string with magnetic flux $\chi/10 - \mathcal{B}$ (see Table \ref{tab:charges}, second last column.)}
\label{fig:E6_chi_2}
\end{center}
\end{figure}

The monopole-string structure associated with the symmetry breaking $E_6 \to SO(10) \to SU(5) \times U(1)_\chi \to SU(5) \times Z_2$ is shown in Fig.~\ref{fig:E6_chi_2}. The string is topologically stable because of the mismatch between the monopole flux $\chi/5$ and the flux $\chi/10$ that the string can carry. Note that the string carries zero modes of the three right handed neutrinos as well as the $N_i$ matter fields.

The breaking of the $SU(5)$ symmetry to the SM gauge symmetry can be carried out in the usual way using the adjoint field of $E_6$. The final breaking to $SU(3)_c \times U(1)_{\rm em} \times Z_2$ is achieved with the $SO(10)$ 10-plet Higgs contained in the 27 dimensional representation of $E_6$. This ensures that the $Z_2$ gauge symmetry stays unbroken and the string is topologically stable. In order to make sure that the phase change of the electroweak doublets from going around the string is an integral multiple of $2\pi$, we should include $-T^3_L/2$ unit of magnetic flux to the string. The combination $\mathcal{B}=T^3_L/2-Y/10$ is the broken generator orthogonal to $Q$. Including the electroweak symmetry breaking the final monopole-string configuration is displayed in Fig.~\ref{fig:E6_chi_2}.

 A crucial thing to note is that phase change of the up type and down type Higgs doublets is $+ 2 \pi$ and $-2\pi$ respectively, which means that the string is superconducting. The up-type quarks and their antiparticles are the right movers, while the down type quarks and charged leptons, together with their antiparticles are the left movers.
Furthermore, the phase change of the spinorial 16-plet of $SO(10)$ matter fields in the 27-plet is $\exp (\pm i\pi)$, which is a reflection of the unbroken $Z_2$ symmetry.
\section{Conclusions}
\label{sec:conc}
We have shown the presence of both metastable and topologically stable superconducting strings in $E_6$ grand unified models. For the metastable strings susceptible to breaking via quantum tunneling of monopole-antimonopole pairs, we ensure that the monopole flux and the flux carried by the string are precisely matched. If appropriately heavy, the metastable strings seem to provide a satisfactory explanation of the stochastic gravitational background observed by the PTA experiments. In the presence of an unbroken gauge $Z_2$ symmetry, the superconducting metastable strings also carry zero modes of a compelling dark matter candidate in $E_6$. The topologically stable string whose presence is guaranteed by the unbroken $Z_2$ symmetry also turns out to be superconducting. Finally, the flux matching condition also ensures that the AB phase change for all matter and Higgs fields in $E_6$, if taken around these metastable strings, is an integral multiple of $2\pi$. The 16 dimensional matter fields in $SO(10)$ acquire an AB phase of $\exp(\pm i\pi)$ if taken around the topologically stable $Z_2$ string.

\section{Acknowledgments}
 We thank Edward Witten for helpful correspondence. R.M. is supported by the Institute for Basic Science under the project code: IBS-R018-D3.
\appendix

\bibliographystyle{JHEP}
\bibliography{sccs}

\end{document}